\documentclass[]{emulateapj}

\usepackage{xcolor}

\def\ms{\hbox{m$\;$s$^{-1}$}}

\shorttitle{Wave heating of partially ionized solar chromosphere}

\shortauthors{Shelyag et al.}

\begin{document}

\title{Heating of the partially ionized solar chromosphere by waves in magnetic structures }

\author{S. Shelyag\altaffilmark{1}, E. Khomenko\altaffilmark{2,3}, A. de Vicente\altaffilmark{2,3}, D. Przybylski\altaffilmark{1}}
\email{shelyag@gmail.com}
\altaffiltext{1}{Department of Mathematics and Information Sciences, Northumbria University, Newcastle upon Tyne, NE1 8ST, UK}
\altaffiltext{2}{Instituto de Astrof\'{\i}sica de Canarias, 38205, C/ V\'{\i}a L{\'a}ctea, s/n, La Laguna, Tenerife, Spain}
\altaffiltext{3}{Departamento de Astrof\'{\i}sica, Universidad de La Laguna, 38205, La Laguna, Tenerife, Spain}

\begin{abstract}
In this paper, we show a ``proof of concept'' of the heating mechanism of the solar chromosphere due to 
wave dissipation caused by the effects of partial ionization. Numerical modeling of non-linear
wave propagation in a magnetic flux tube, embedded in the solar
atmosphere, is performed by solving a system of single-fluid quasi-MHD
equations, which take into account the ambipolar term from the generalized Ohm's
law. It is shown that perturbations caused by magnetic waves can be effectively
dissipated due to ambipolar diffusion. The energy input by this mechanism is
continuous and shown to be more efficient than dissipation of static
currents, ultimately leading to chromospheric temperature increase in
magnetic structures. 
\end{abstract}


\section{Introduction}

Understanding the physical mechanisms leading to heating of the solar
atmosphere is one of the primary questions in solar physics and a long-standing
puzzle. In recent years, the importance of partial ionization for the
processes in solar plasma is becoming clear. It has been shown
that the current dissipation enhanced in the presence of neutrals in
the plasma not entirely coupled by collisions can be essential for the energy
balance of the chromosphere \citep{DePontieu1998, Judge2008,
  Krasnoselskikh2010, MartinezSykora+etal2012, Khomenko+Collados2012}. In the
latter paper it was shown how the dissipation of static currents in
non-current-free solar magnetic flux tubes allows to release a large amount of
energy into the chromosphere, potentially able to compensate its radiative
losses, \citep{Khomenko+Collados2012b}.

Solar atmosphere is far from stationary. It is filled with
waves and rapidly varying flows, and pierced by dynamic magnetic 
structures, originating in the solar interior. As it has been demonstrated
\citep{Osterbrock1961, goodman2005, steiner2008, goodman2010, Krasnoselskikh2010, goodman2011, abbett2012, Shelyag+etal2012, Shelyag+etal2013}, the magnetized plasma motions in the
solar photosphere create an electromagnetic (Poynting) energy flux sufficient to
heat the upper solar atmosphere. A significant part of this flux is
produced in the form of Alfv{\'e}n waves
\citep{Shelyag+Przybylski2014}. These waves are notoriously difficult to
dissipate if an ideal plasma assumption is employed.

In the present study we extend the analysis by \citet{Khomenko+Collados2012}
including time-dependent oscillatory perturbations. Magnetic waves (fast
magneto-acoustic and Alfv\'en waves) produce perturbations in
the magnetic field. These perturbations create currents at all times
and locations in the magnetised solar atmosphere. Similarly to static currents,
the currents produced by waves in a partially ionized magnetised chromosphere
can be efficiently dissipated due to the presence of ambipolar diffusion. This
would provide a constant and efficient influx of energy into the chromospheric
layers.

It was long known that the presence of neutral atoms in partially ionized
plasmas significantly affects excitation and propagation of waves
\citep{Kumar+Roberts2003, Khodachenko2004, Forteza2007,
  Pandey2008, Vranjes2008, Soler2009, Zaqarashvili2011}. The relative
motion between the ionized and charged species causes an increase of collisional
damping of MHD waves in the photosphere, chromosphere and prominence plasmas, in
a similar way that any electro-magnetic wave is dissipated in a
collisional medium. These mechanisms were investigated for high-frequency waves
and were shown to be important for frequencies close to the collisional
frequency between the different species \citep{Khodachenko2004, Forteza2007, Pandey2008}. In the cited
works the damping was analyzed in terms of the damping time of
waves. Here we study the energy input due to this damping.

\section{Equations and numerical method}

In the solar chromosphere the collisional coupling of the plasma is
still strong \citep{Zaqarashvili2011, Khomenko+etal2014}. In this case, it is convenient to use a single-fluid quasi-MHD approach i.e, including non-ideal terms not taken into account by the ideal MHD approximation instead of solving full multi-fluid equations. The conservation equations for the multi-species solar plasma with scalar pressure and no heat flux can be written as follows:
\begin{equation} 
\frac{\partial \rho}{\partial t} + \vec{\nabla}\cdot\left(\rho\vec{u}\right) =  0,
\end{equation}
\begin{equation} 
\frac{\partial (\rho\vec{u})}{\partial t} + \vec{\nabla}\cdot(\rho\vec{u} \otimes \vec{u}) + \nabla p  = \vec{J}\times\vec{B} + \rho\vec{g}  ,
\end{equation}
\begin{equation}
\label{eq:energy-single-p}
\frac{\partial e}{\partial t} + \vec{\nabla}\cdot( \vec{u}e) +p\nabla\cdot \vec{u}  = \vec{J}  \cdot [\vec{E} + \vec{u} \times \vec{B}], 
\end{equation}
where the internal energy $e$ is computed from the scalar pressure according to the ideal equation of state:
\begin{equation}
e=p/(\gamma-1).
\end{equation}

The rest of the notation is standard \citep[for details see][]{Khomenko+etal2014}. This system of equations is closed by the generalized Ohm's law, where we only include Ohmic and ambipolar terms:
\begin{equation} 
[\vec{E} + \vec{u}\times{\vec{B}}] =  \eta\vec{J}  - \eta_A\frac{[(\vec{J} \times \vec{B}) \times \vec{B}]}{|B|^2} 
\end{equation}
with resistivity coefficients (in units of $ML^3/TQ^2$) defined as
\begin{equation} 
\eta=\frac{\rho_e}{(en_e)^2}\left(\sum_{a} \nu_{e; ai} + \sum_{b} \nu_{e; bn} \right),
\end{equation}
\begin{equation}
\eta_A=\frac{\xi_n^2 |B|^2}{\alpha_n},
\end{equation}
where
\begin{equation}
\alpha_n= \sum_b\rho_e\nu_{e; bn} + \sum_{a}\sum_b\rho_{a}\nu_{ai; bn}.
\end{equation}
In these equations the inter-species collision frequencies $\nu$ are those between electrons and ions of the specie $a$ ($\nu_{e; ai}$), between electrons and neutrals of the specie $b$ ($\nu_{e; bn}$) and those between ions and neutrals of different species ($\nu_{ai; bn}$). These frequencies, together with the densities $\rho$, the electron number density $n_e$, and the neutral fraction $\xi_n$ are calculated according to \citet{Khomenko+Collados2012, Khomenko+etal2014}. We do not include the rest of the terms in the generalized Ohm's law as these terms either do not contribute to the energy balance, as Hall term, or can be considered small for the solar atmospheric conditions \citep{Khomenko+etal2014}. Nevertheless, it should be noted that the battery term can potentially increase the current and, therefore, the dissipation, as it was demonstrated by \citet{diaz2014}.

Finally, the induction equation reads as:
\begin{equation} 
\frac{\partial\vec{B}}{\partial t}  =  \vec{\nabla}\times \left[(\vec{u}\times\vec{B})  - \eta\vec{J}  + \eta_A\frac{[(\vec{J} \times \vec{B}) \times \vec{B}]}{|B|^2}  \right].
\end{equation}

The physical magnitude of $\eta$ is too small to produce a measurable effect. We used a numerical analog of the $\eta$, denoted as $\eta_{\rm hyp}$ (hyperdiffusion). The value of $\eta_{\rm hyp}$ is allowed to vary in space and time according to the structures developed in the simulations, so that it reaches larger values at places where the variations are small-scale \citep{Vogler2005, Felipe+etal2010}. $\eta_{\rm hyp}$, together with the grid resolution, defines the maximum Reynolds number of the simulations. The effective magnetic Reynolds number due to $\eta_{\rm hyp}$ in the simulation  presented in this paper (as computed directly from the derivatives of physical quantities entering its definition) varies between $10^0$ and $10^3$, with the smallest values localised in narrow layers at the regions where the gradients are strongest, i.e. at the shocks. $\eta_A$, however, can reach values significantly larger than $\eta_{\rm hyp}$ decreasing the effective {\it physical} magnetic Reynolds number to $10^{-1}-10^0$ in the layers above 1200 km, where $\eta_A$ is large.

The equations above are solved by means of the code {\sc  mancha} \citep{Khomenko+etal2008, Felipe+etal2010, Khomenko+Collados2012}. We evolve $n_e$ in time via the Saha equation. The simulations are done
in full 3D. We use 10 grid points Perfectly Matched Layer \citep[PML,][]{berenger1994} boundary conditions on the
side and top boundaries. At the bottom boundary a time-dependent condition for
all variables is provided to drive waves into the domain.

\begin{figure}
\includegraphics[width=8cm]{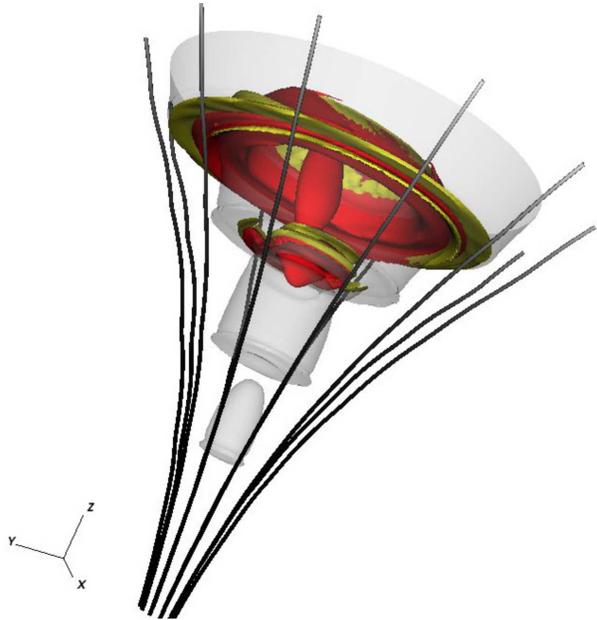}
\caption{Geometry of the magnetic field. 3D view combining the magnetic field
  lines (black), location of strong currents (yellow), relative temperature
  enhancements ($+500~\rm{K}$, red), and plasma $\beta =1$ sufraces (grey) at the
  simulation time t=240 sec.}
\label{fig:3dview}
\end{figure}

\begin{figure*}[!ht]
\center
\includegraphics[width=5cm]{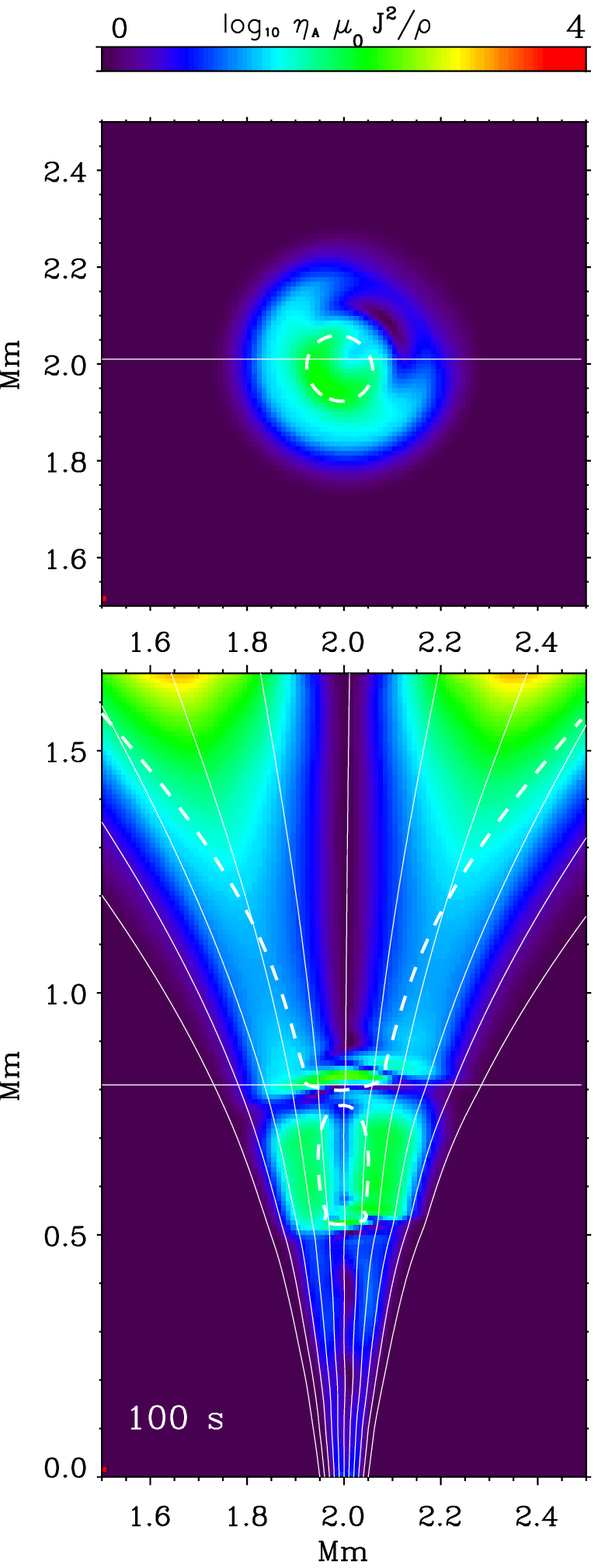}
\includegraphics[width=5cm]{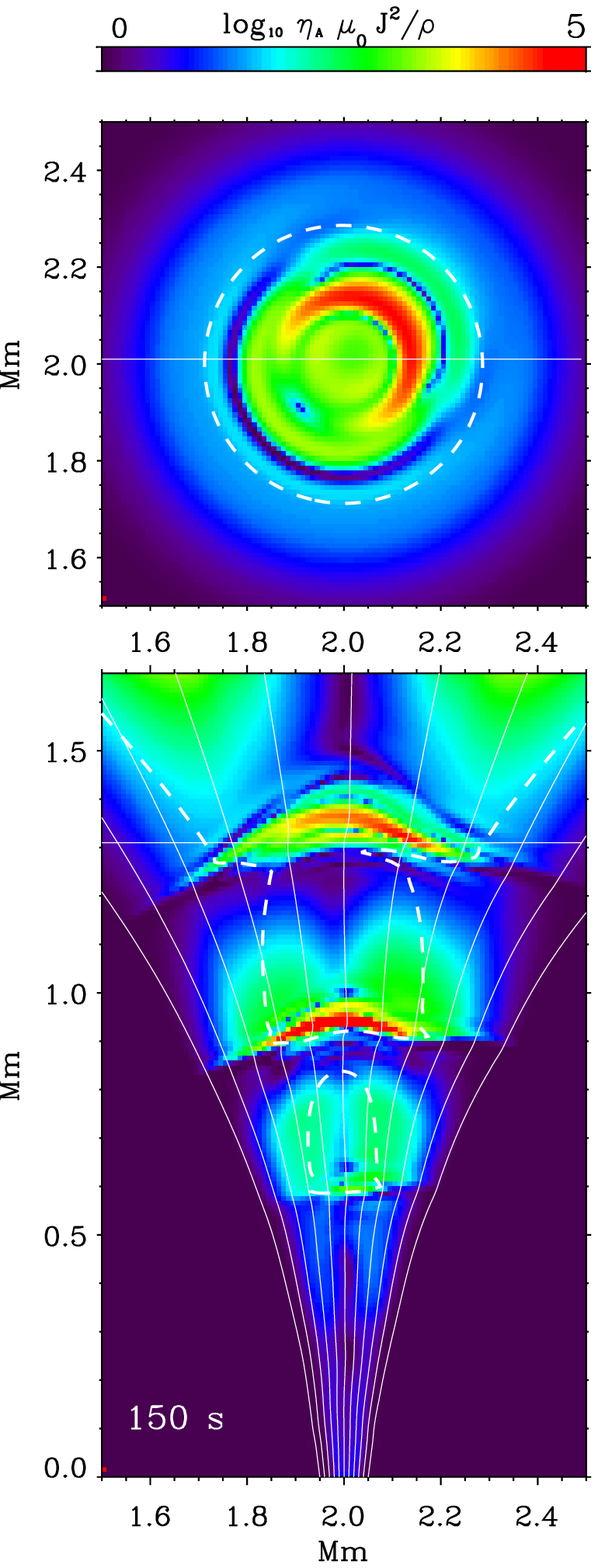}
\includegraphics[width=5cm]{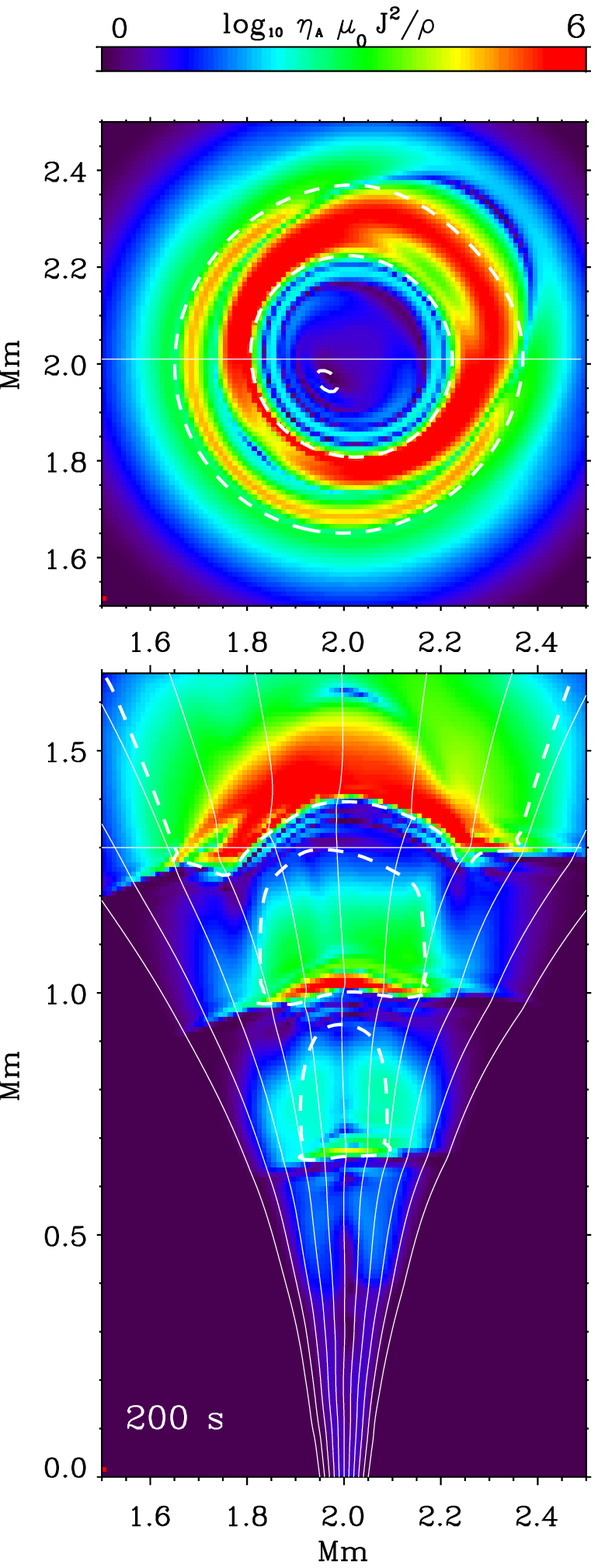}
\caption{Temporal evolution of the ambipolar heating term $\log \eta_AJ_{\perp}^2\mu_0/\rho$. The horizontal and vertical cuts through the domain are shown in the top and bottom panels, respectively, for 100, 150 and 200 seconds of simulated physical time. The positions of the cuts are marked by thick white lines. The thin white lines are magnetic field lines. The dashed lines correspond to plasma $\beta=1$ contours in the perturbed models.}
\label{fig:ambiheating}
\end{figure*}

\section{Flux tube model}

A single 3D Schl\"uter-Temesvary-like axisymmetric self-similar magnetic flux
tube, embedded into the VALIIIC solar atmospheric model
  \citep{valc}, similar to the one in \citet{Fedun+etal2011a,
  Fedun+etal2011b, Gent+etal2013}, is used as the initial equilibrium
model in the simulation (see Fig.~\ref{fig:3dview}). The components of the magnetic field are calculated as follows:
\begin{eqnarray}
B_x = -x B_{0}(z) G \frac{d B_0(z)}{d z},\\
B_y = -y B_{0}(z) G \frac{d B_0(z)}{d z},\\
B_z = B_{0}^2(z) G,
\label{b_eq}
\end{eqnarray}
where $G$ is a piecewise-parabolic function with the dimension of $B^{-1}$,
describing the opening of the magnetic flux tube. $B_{0}(z)$ is chosen
so that it decreases with height, similar to the gas
  pressure. This choice allows to avoid negative values of gas
pressure within the magnetic flux tube, while keeping plasma-$\beta$ low. The
magnetic field components were used to recalculate (using the
  magnetohydrostatic equation) the pressure and density to obtain a 3D
  magnetohydrostatic configuration. The
simulation box is set to have physical dimensions of 4$\times$4$\times$1.84
$\rm{Mm^3}$, with a uniform resolution of 10 km. The bottom boundary of the simulation box is located in the
deep photosphere. The parameters of the magnetic field are chosen such that the
magnetic field strength at the axis of the magnetic flux tube is
  $1.4~\rm{kG}$ at the bottom ($z=0$) and $10~\rm{G}$ at the top
  ($z=1.84~\rm{Mm}$) of the domain. This leads to the plasma $\beta=1$ height of
  about $0.7~\rm{Mm}$ at the axis of the tube.

Self-consistent periodic perturbations of the velocity vector, pressure, and
density are introduced according to \cite{Mihalas+Mihalas1984}. For the
simulations described here we used the driving frequency of $\nu=\omega/2\pi=25$
mHz. The choice of the driving frequency was motivated by the
results of \citet{Shelyag+etal2013} who showed that Alfv\'en waves with such
periods can be generated in intergranular magnetic flux tubes by turbulent
convective motions. To break the symmetry, the acoustic pulse was placed at
$100~\rm{km}$ off the flux tube axis in the field-free atmosphere and was
limited in horizontal extent by a Gaussian of $100~\rm{km}$ FWHM. The
perturbation amplitude was set to 500 \ms. The duration of the simulation was $350~\rm{s}$.

\begin{figure*}
\center
\includegraphics[width=12cm]{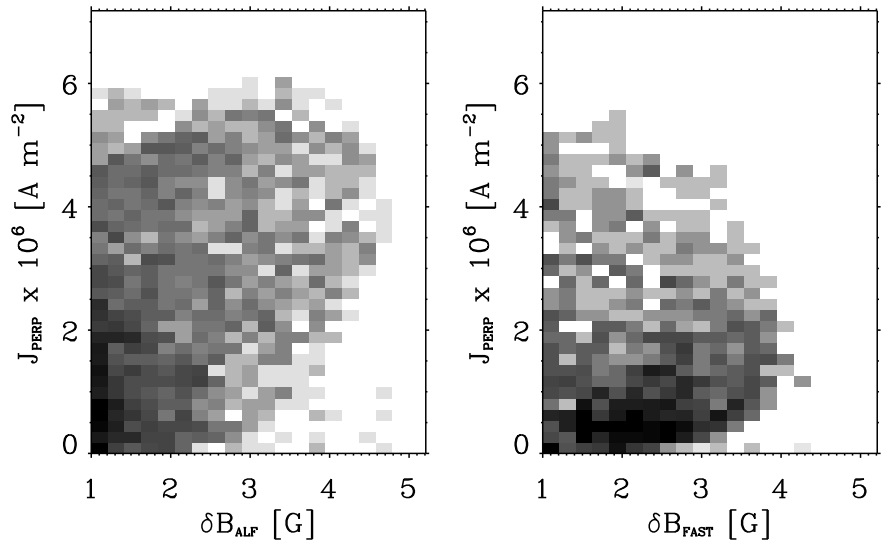}
\caption{Occurrence plots of the perpendicular current $J_\perp$ vs the amplitude of the magnetic field fluctuation due to the Alfv\'en wave (left) and the fast wave (right). The magnetic field fluctuations were obtained using the projections of the magnetic field perturbations onto the directions, given by Eqs.~\ref{eq:projections1}-\ref{eq:projections3}.}
\label{fig:core}
\end{figure*}

In order to understand the effect of plasma oscillations on chromospheric
heating, we perform three simulation runs with exactly the same numerical
parameters, but different driving. In the first run (denoted as $AD$ below), we
include the ambipolar term as a perturbation. This case is analogous to
\citet{Khomenko+Collados2012}. In the second run (denoted as $W$), we include only the wave
driving, with no ambipolar term. In the third run (denoted as $ADW$), we include both the
  wave driving and the ambipolar diffusion term. In the $AD$ run, the
ambipolar term alone acting on a static flux tube produces dissipation of its
static currents, which creates perturbations in all thermodynamical
parameters and, in particular, in temperature. In the $ADW$ case, both wave and ambipolar drivings produce variations of temperature and therefore it is difficult to separate the effects of waves from the effect of
static current dissipation. For that we perform a flatfield-like procedure. In
order to determine the heating caused by the current dissipation of the currents
produced only by the waves, we subtract from $ADW$ case the heating produced in the
$AD$ case, as well as the variations due to waves produced in the $W$ case.

\section{Wave propagation and heating}

The acoustic pulse set in the field-free atmosphere generates essentially
acoustic (fast) waves. These waves gradually enter the flux tube, cross
the $c_s=v_a$ level and generate fast and slow magneto-acoustic waves
and Alfv\'en waves via the mode transformation. The latter waves propagate in
the magnetic flux tube and create perpendicular currents, which then
dissipate due to ion-neutral friction (see Fig.~\ref{fig:3dview}). This process is demonstrated in
Fig.~\ref{fig:ambiheating}, where the logarithm of the ambipolar diffusion
heating term $\log \eta_A J^2 \mu_0 / \rho$ is shown. In the first
column, the heating term is shown for $100~\rm{s}$ of  simulation time. At this time, the perturbation has not reached the upper layers of the domain above $1~\rm{Mm}$. The heating term values are below $10^4~\rm{J~kg^{-1} s^{-1}}$, with the location of the largest values at the top of the domain. The values of $\eta_A$ reach their maximum in the interior of the flux tube, while the static currents are largest at the flux tube boundaries. As a consequence, the heating by dissipation of static currents is strongest at the tube walls. Note that the wall heating is present at all heights, but is largest in the upper part of the domain, as a consequence of the strong increase of $\eta_A$ with height \citep[see similar case in][]{Khomenko+Collados2012}.
At $150~\rm{s}$ the situation changes. The wave perturbations reach the upper part of the domain, and the maximum of the heating term is localised in the regions around the chromospheric shocks. Further on, at $200~\rm{s}$ (third column of Fig.~\ref{fig:ambiheating}), the ambipolar diffusion heating rate reaches $10^6~\rm{J~kg^{-1} s^{-1}}$, which is two orders of magnitude higher than for the case of dissipation of static currents (first column of the figure). The high heating rate region is now not confined to the chromospheric shock anymore, and occupies the whole upper part of the domain above $\sim 1.2~\rm{Mm}$ height.

\begin{figure*}
\center
\includegraphics[width=16cm]{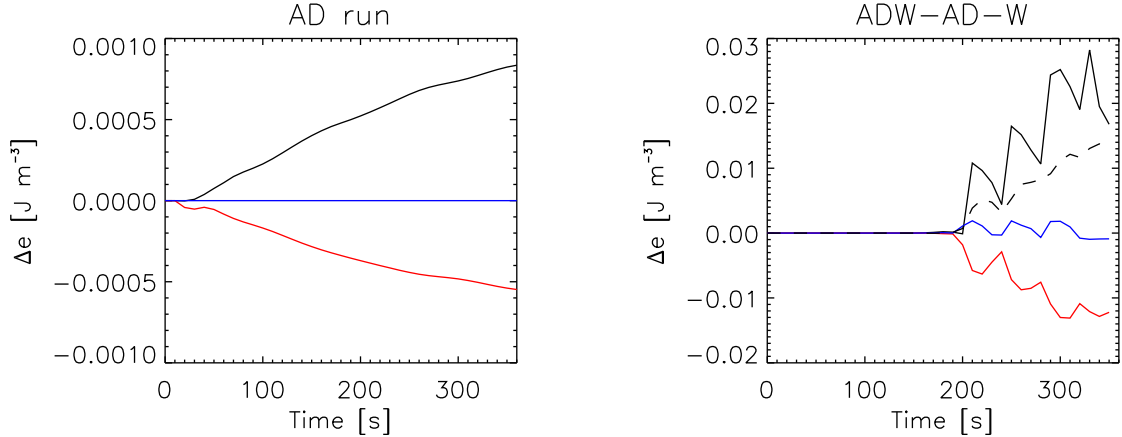}
\caption{Conversion into thermal energy of the static currents and the currents induced by oscillatory perturbation in the domain. Left panel: evolution of the kinetic (blue), magnetic (red) and thermal (black) energies averaged over the volume of the numerical domain bounded by the surface $\beta < 1$ within the flux tube for the $AD$ simulation. The initial thermal and magnetic energy values have been subtracted. Right panel: evolution of the difference between the energies in the $ADW$ simulation and $AD$ and $W$ simulations. Dashed curve corresponds to the absolute value of the sum of the magnetic and kinetic energy differences.}
\label{fig:energy-in}
\end{figure*}

In order to separate the different modes in the atmosphere where $v_a > c_s$, and study their behavior and relation to heating, we projected the velocities in the computational domain onto the three characteristic directions, defined by the following vectors \citep{khomenko+cally2012, Cally+Khomenko2015}:
\begin{equation}
 \hat{e}_{\rm long} = [\cos\phi \sin\theta, \, \sin\phi \sin\theta, \, \cos\theta]; 
\label{eq:projections1}
\end{equation}
\begin{equation}
\hat{e}_{||} = [ \sin\phi, \, -\cos\phi,  0];
\label{eq:projections2}
\end{equation}
\begin{equation}
\hat{e}_{\perp} = [\cos\phi \cos\theta, \, \sin\phi \cos\theta, \, -\sin\theta].
\label{eq:projections3}
\end{equation}
Here, $\theta$ is the inclination and $\phi$ is the azimuth of the magnetic field. 
The first vector $\hat{e}_{\rm long}$ is directed along the magnetic field
lines. The second vector $\hat{e}_{||}$ gives the direction perpendicular to
$\vec{B}$ in the plane containing the field lines and gravity. The third vector is perpendicular to the first two. These
projections allow to separate the fast wave in the $\hat{e}_{\perp}$ projection
and the Alfv\'en wave in the $\hat{e}_{||}$ projection. This was used to analyse
the relative influence of slow and Alfv\'en waves on the perpendicular current, and, subsequently, on the ambipolar heating in the simulation. Fig.~\ref{fig:core} shows the occurrence plots of the perpendicular current $J_\perp$ versus the magnetic field fluctuations along $\hat{e}_{||}$ (Alfv\'en wave, left panel), and along $\hat{e}_{\perp}$ (fast wave, right panel). Clearly, Alfv\'en waves have a stronger effect on the chromospheric non-ideal plasma heating. Furthermore, the perpendicular current increases with the increase of Alfv\'en wave perturbation in the magnetic field, while the occurrence plot shows no significant dependence of the perpendicular current on the fast wave magnetic field perturbation amplitude.

Slow waves, aligned with $\hat{e}_{\rm long}$, are also produced by the oscillatory source in our simulations. However, they do not perturb the magnetic field in the perpendicular to the field line direction, therefore have only a minor effect on the energy balance in the modelled chromosphere.

Through the heating term in the energy equation Eq.~\ref{eq:energy-single-p} associated with $\eta_A$, the wave energy is converted into thermal energy. This process is demonstrated in Fig.~\ref{fig:energy-in}. In this plot, the time dependences of the volume-averaged thermal (black curve), kinetic (blue curve) and magnetic (red curve) energy densities per unit volume are shown. The averaging of the energy density values is carried over the volume around the tube axis, extending in the simulated chromosphere between $1.3$ and $1.65~\rm{Mm}$. The volume boundaries lie within the plasma $\beta < 1$ surface, which also coincides with the region where $\eta_A$ is maximal. In the left panel of this figure, it is straightforward to see that the thermal energy density in the $AD$ simulation increases over time due to the dissipation of static currents in the initial non-current-free magnetic field configuration. Correspondingly, the magnetic field energy density decreases over time.


The right panel of Fig.~\ref{fig:energy-in} shows the difference between the corresponding energy densities in the $ADW$ simulation and $W$ simulation. The energy densities from the $AD$ run (left column of Fig.~\ref{fig:energy-in}) were also subtracted. It is evident from the plot that the dissipation of the currents induced by the oscillatory source leads to the decrease of the kinetic and magnetic energy density, as the magnetic energy gain due to the source action is the same in the ambipolar and ideal simulations, while the additional dissipation present in the ambipolar simulation decreases the magnetic field energy and converts it into the thermal energy. The average thermal energy per unit volume being deposited to the low-$\beta$ chromospheric plasma due to wave dissipation is about 10-20 times greater than the one deposited due to the dissipation of static currents. 

The computed spatially-averaged thermal energy density difference due to dissipation of the perpendicular currents caused by waves corresponds to the additional temporally-averaged thermal energy flux of $\sim 100~\rm{J~m^{-2}~s^{-1}}$ through the magnetised chromosphere, assuming uniform heating distribution for simplicity (similar to \citet{ballegooijen2011}). While this estimate is about one order of magnitude less than generally accepted value required to compensate the radiative energy losses \citep[$2\div 4\cdot 10^3~\rm{J~m^{-2}~s^{-1}}$,][]{Osterbrock1961,Shelyag+etal2012,Arber+etal2015}, the Poynting flux in our simulations at the chromospheric level ($z=0-0.5~\rm{Mm}$) is equal to $\sim 700~\rm{J~m^{-2}~s^{-1}}$, which is also nearly an order of magnitude less than that value. Therefore, our result suggests that waves, produced by photospheric motions, are an important energy source to provide heating of the solar chromosphere.

\begin{figure*}
\center
\includegraphics[width=16cm]{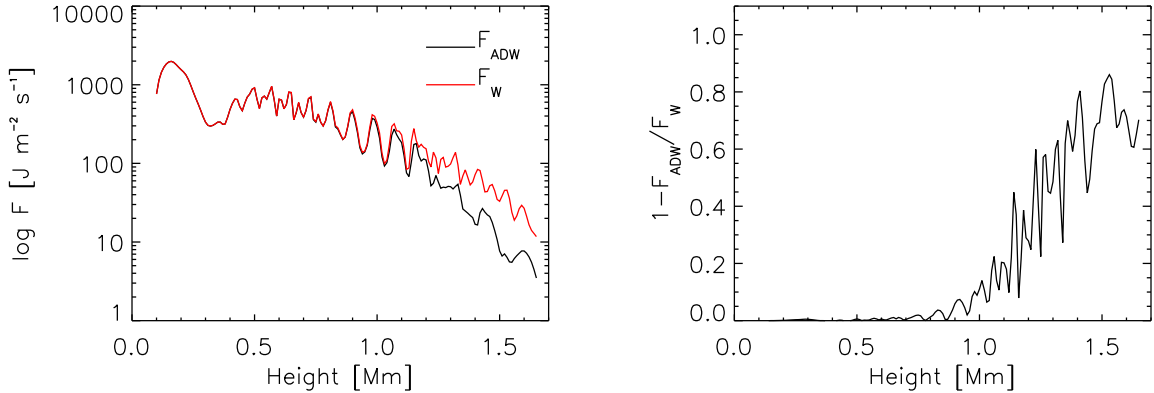}
\caption{Poynting flux absorption. Left panel: height dependence of temporally and horizontally averaged Poynting flux for the $W$ simulation  ($F_W$, red curve) and for the $ADW$ simulation with the wave source and ambipolar diffusion term ($F_{ADW}$, black curve). Right panel: dependence of Poynting flux absorption $1-F_{ADW}/F_W$ on height in the simulation.}
\label{fig:pf-ratio}
\end{figure*}

Further evidence is provided by the measurements of Poynting flux absorption. The left panel of Fig.~\ref{fig:pf-ratio} shows horizontally-averaged Poynting flux in the simulations with the oscillatory source and with (black curves) and without (red curves) ambipolar diffusion. The averaging is carried out over a narrow region within $200~\mathrm{km}$ range around the tube axis. The Poynting flux in the $W$ simulation decreases with height due to the tube expansion out of the averaging region boundaries, and does not behave differently from the $ADW$ simulation in the lower solar atmosphere. Starting from $1.1~\rm{Mm}$ height, the ideal and non-ideal simulation curves diverge. The decrease in the Poynting flux by the factor of up to 0.8 is produced by ambipolar diffusion only. The Poynting flux absorption dependence on height is shown in the right panel of Fig.~\ref{fig:pf-ratio}. The spatially averaged absorption coefficient is $d(1-F_{ADW}/F_{W})/dz = 0.47~\rm{Mm^{-1}}$ for the tube axis region.

\section{Conclusions and discussion}
In this paper we investigated chromospheric heating by oscillations in the chromospheric magnetic fields. An initial study was performed using a single magnetic flux tube, rooted in the photosphere in magneto-hydrostatic equilibrium. A monochromatic oscillatory perturbation was placed off the tube axis in the deep photosphere to generate magneto-acoustic oscillations in the magnetic flux tube. The currents, corresponding to fast magneto-acoustic and Alfv\'en waves, have been shown to be dissipated by the ambipolar diffusion mechanism in the simulated chromosphere. It has been shown that for a source amplitude of $500~\rm{m~s^{-1}}$, which mimicks photospheric motions, the heating due to the perturbation is two orders of magnitude larger than that due to the static currents dissipation. The perturbation we used generates all types of MHD waves in the magnetic flux tube. It has been shown, however, that the Alfv\'enic perturbation component has a stronger effect on the heating.

Generally, it is expected that oscillations with higher ($1-10~\rm{Hz}$)
frequencies are more efficiently dissipated into heat compared to lower
(similar to the ones used in this paper) frequency oscillations \citep[e.g.][]{Soler+etal2015, Arber+etal2015}. On the other hand, a power law is expected for the velocity power spectrum in the solar photosphere, including its high-frequency part, currently not observed \citep{Goldreich+etal1994, Musielak+etal1994, Stein+Nordlund2001, Fossum+Carlsson2005}, leading to less power at higher frequencies. Consequently, it is currently difficult to draw a unique conclusion on a sole mechanism for the chromospheric heating based on a single simulation. Nevertheless, this study should be understood as an initial proof of concept of the chromospheric energy balance based on MHD wave absorption in non-uniform three-dimensional magnetised chromospheric plasmas, with a more detailed study of the absorption on the source frequency, amplitude and simulation resolution to follow. 

Furthermore, as the simulation has to resolve the smallest spatial and temporal scales in the system, which in the case of frequencies of the order of $10~\rm{Hz}$ is defined by the oscillation wavelength and period, an extremely high spatial resolution of the order of few hundred meters would be necessary to directly test high-frequency currents dissipation. Therefore, even with currently available computational resources some extrapolation would be required.

\acknowledgments 
EK was partially supported by the Spanish Ministry of Science through projects AYA2010-18029, AYA2011-24808, and AYA2014-55078-P. This work contributes to the deliverables identified in FP7 European Research Council grant agreement 277829, ``Magnetic connectivity through the Solar Partially Ionized Atmosphere''. This research was supported by the SOLARNET project, funded by the European Commission's FP7 Capacities Programme under the Grant Agreement 312495. This research was undertaken with the assistance of BSC-CNS MareNostrum facilities, the resources provided at the Centre for Astrophysics \& Supercomputing of Swinburne University of Technology (Australia), NCI National Facility systems at the Australian National University through the National Computational Merit Allocation Scheme supported by the Australian Government, and at the Multi-modal Australian ScienceS Imaging and Visualisation Environment (MASSIVE).

\providecommand{\noopsort}[1]{}\providecommand{\singleletter}[1]{#1}%

\end{document}